# XML REWRITING ATTACKS: EXISTING SOLUTIONS AND THEIR LIMITATIONS

Azzedine Benameur
*SAP Research-Security & Trust*
*Mougins, FRANCE*

Faisal Abdul Kadir
*SAP Research-Security & Trust*
*Mougins, FRANCE*

Serge Fenet
*LIRIS CNRS UMR 5205,University, Lyon 1*
*Lyon, FRANCE*

**ABSTRACT**

Web Services are web-based applications made available for web users or remote Web-based programs. In order to promote interoperability, they publish their interfaces in the so-called WSDL file and allow remote call over the network. Although Web Services can be used in different ways, the industry standard is the Service Oriented Architecture Web Services that doesn't rely on the implementation details. In this architecture, communication is performed through XML-based messages called SOAP messages. However, those messages are prone to attacks that can lead to code injection, unauthorized accesses, identity theft, etc. This type of attacks, called XML Rewriting Attacks, are all based on unauthorized, yet possible, modifications of SOAP messages. We present in this paper an explanation of this kind of attack, review the existing solutions, and show their limitations. We also propose some ideas to secure SOAP messages, as well as implementation ideas.

## 1. INTRODUCTION

### 1.1 Context

Web Services are web-based applications made available for web users or remote Web-based programs. In order to promote interoperability, they publish their interfaces and allow remote call over the network. These interfaces are published with a precise syntax in a file called the Web Specification Description Language file [14]. Web Services can be used in different architectures: Remote Procedure Call (RPC), Service Oriented Architecture (SOA) or Representational State Transfer (REST). The industry standard is currently SOA Web Services, because it is independent from the underlying implementation. In SOA Web Services, communications are made through XML-Based messages called Simple Object Access Protocol (SOAP) [2] messages. Unfortunately, those messages are prone to attacks that can lead to several consequences such as unauthorized access, disclosure of information, identity theft... These attacks are all based on an on-the-fly modification of SOAP messages, referred as XML rewriting Attacks [9].

### 1.2 Motivations

This security issue has already been addressed by several mechanisms. For example, WS-Policy [12] and WS-Security [13] are both part of the web service security stack. They provide mechanisms to ensure end-to-end security and allow one to protect some sensitive parts of a SOAP message by the mean of XML Digital Signature [3]. However, it as been proven that naïve use of the XML Digital Signature and WS-Security could let an attacker modify SOAP messages without altering the signature [4, 9]. Therefore, efforts have been made to protect the structure of the SOAP message itself [5, 6]. While this approach seems to be the best way towards protection of SOAP messages it still has some flaws. This paper aims at presenting existing solutions to detect XML rewriting attacks [5, 6, 9, 4] and show their limitations. We exhibit examples showing that these solutions can fail, and present some ideas to improve the detection of these attacks. The paper is structured as follows. Section 2 presents the weakness on which this attack is based, and shows an example. Section 3 presents existing solutions. Section 4 emphasizes their limitations. Section 4 presents our ideas, as well as possible implementations and finally section 5 concludes.

## 2. ATTACK SCENARIO

## 2.1 Signature's Weakness

```
<Signature ID?>
    <SignedInfo>
      <CanonicalizationMethod/>
      <SignatureMethod/>
      (<Reference URI? >
        (<Transforms>)?
        <DigestMethod>
        <DigestValue>
        </Reference>)+
    </SignedInfo>
     <SignatureValue>
    (<KeyInfo>)?
    (<Object ID?>)*
    </Signature>
```

**Figure 1. XML Digital Signature Structure**

Parts of a SOAP message can be encrypted and/or signed using XML Digital Signature [3]. The signature is used by the receiver to check integrity of the message and authenticity of the sender. The structure of the signature is depicted in Figure 1. We can see that the signed data object is referenced using a reference URI within the XML Signature element, that is a child of the XML Signature element. Thus the signed object is inside the XML Signature Element. The signed data object contains the XML Signature Element, which contains its signature, within it. Therefore the signed data object is the parent of its signature element. The weakness of the signature lies within its verification. Verification of the XML digital signature is done in two steps:

➤ Reference Validation:

In this step the digest value for each referenced data object is checked for validity. First the *<SignedInfo>* element is canonicalized using the *CanonicalizationMethod* specified under the *<SignedInfo>* element. Then, for each *Reference* element, the referenced data object is retrieved and a digest value is calculated on that data object using the digest method specified under the *<Reference>* element. The resulting digest value is compared with the digest value specified under the *<Reference>* element. If these two values are same then the verification proceeds to the second reference element . Otherwise it generates an error message.

➤ Signature Validation:

If the Reference Validation step passed successfully then comes the Signature Validation step. In this step the keying information, specified in the *<KeyInfo>* element of the *<Signature>* element is retrieved possibly from an external source. Then the Signature method is determined from the *<SignatureMethod>* element of the *<SignedInfo>* element. The Keying Information and *SignatureMethod* , are used to validate the Signature value specified under the *<SignatureValue>* element of the *<Signature>* element.

XML Digital Signature allows non-contiguous objects of an XML dataset to be signed separately. The signed object may be referenced using an indirection (URI) by the *Reference* element of the Signature. This indirect referencing does not give any information regarding the actual location of the signed object. Therefore, the signed object can easily be relocated and the Signature value still remains valid. In cases where the location of the data object is important in the interpretation of the semantics associated with the data, this can be exploited by an adversary to gain unauthorized access to protected resources [9]. This is the main limitation of XML Digital Signature.

## 2.2 Rewriting Example

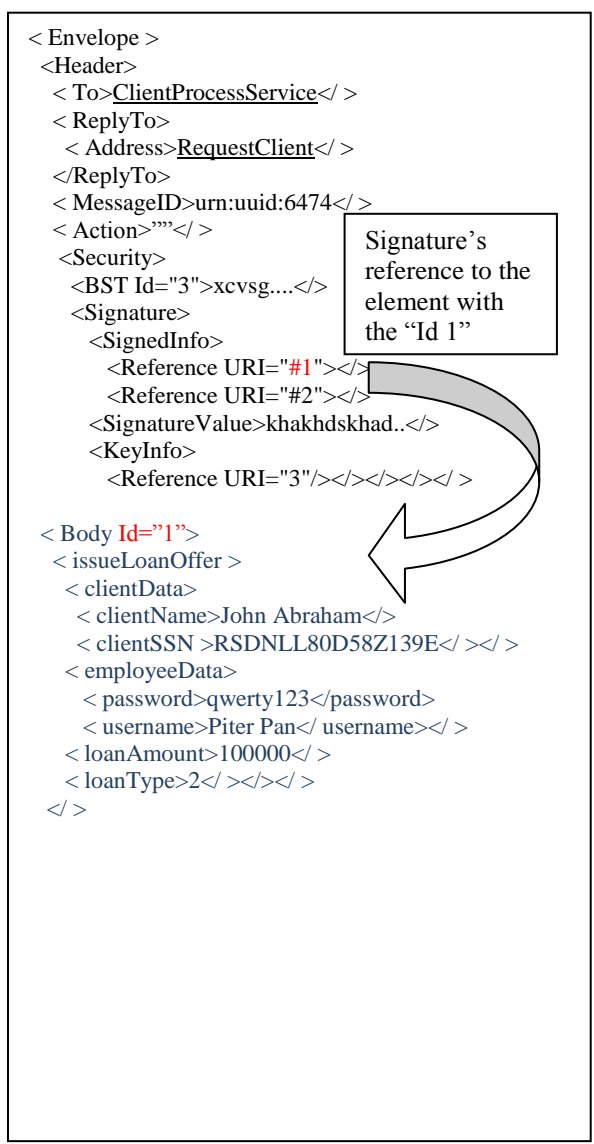

**Figure 2. SOAP Message**

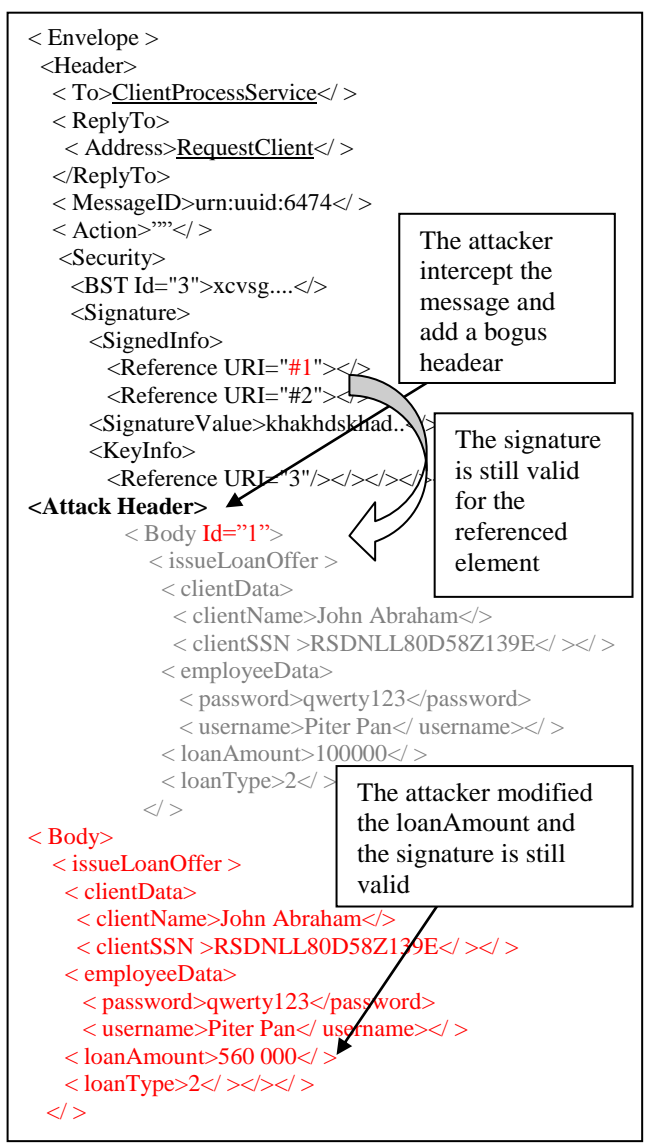

**Figure 3. SOAP message after XML rewriting attack**

We previously introduced the XML Signature weakness. Figure 3 shows an example that demonstrates the ease of modifying SOAP messages without altering the signature. Figure 2 shows a sample SOAP message from banking scenario, this message contains sensitive information such as the client name, the employee name and password along with the loan amount.

The XML rewriting Attack, depicted in Figure 2, is quite straightforward. The attacker encapsulates the *Body* into a bogus tag, then rewrites its own *Body*. When the server receives this message, the validity of the referenced element "*Id 1*" is still valid so the tampering is not detected. In the case of a charge per request service, this attack needs to be taken seriously. Figure 3 presents a simple example, but we could think of a rewriting attack to include code injection, like SQL injection. The latter would allow an attacker to achieve identity theft by rewriting the client name for example, gain access to unauthorized information, using SQL injection to bypass password, and all this without breaking the crypto-mechanism of the Signature.

# 3. RELATED WORKS

Several mechanisms have been previously proposed in order to secure WS communications. We, however, show that these proposals are not sufficient to detect all type of XML rewriting attack.

## 3.1 SOAP Account

In [5, 6] the authors proposed an inline approach that takes into account information about the structure of the SOAP message by adding a new header element called SOAP Account. The SOAP account header contains the following information regarding the structure of a SOAP message:

- ➢ The number Of Child Elements of Envelope.
- ➢ The number Of Header Elements in the SOAP message.
- ➢ The number Of References in each signature.
- ➢ Successor and Predecessor Relationship of Each Signed Object: Parent Element and Sibling Elements.
- ➢ A Possible extension for future improvement.

This SOAP Account element must be signed by the creator using its X.509 certificate. Each successive SOAP node must sign its own SOAP account concatenated with the signature of the previous node.

## 3.2 WS-Policy

Web Service Security Policy [12], if used correctly, can prevent such attacks. However, it is quite difficult to specify all possible security requirements in the WS-Security policy file. The author and the implementer of the security policy needs to be very careful in writing and implementing the policy. For example, in [4], the authors have shown different types of rewriting attacks and the associated policy files for their detection. They have also shown how an attacker can take advantage of a security policy hole in order to get unauthorized access to system resources. According to [4], the semantic of XML elements depends on their location, but in practice XML Signature provides referencing of an element independently of its location. The authors further showed that the flexibility provided by SOAP header can be exploited by an attacker in a naïve way.

## 3.3 WSE Policy Advisor

In [9], the authors proposed a rule-based tool for detecting typical errors in Web Service Enhancements [11] (WSE) configuration and policy. This tool, called Advisor, takes the policy and configuration files of WSE, runs several static queries and generates security reports as well as remedial actions for security flaws. This tool has more than 30 queries that check for syntactic conditions in the policy files. These syntactic conditions are determined by security reviews of the policy and configuration files of WSE. If these security conditions are not met by the policy files, the tool generates a report stating the threat that might occur due to these missing syntactic conditions. It also generates a remedial action that could be used by the author of the policy files to fix the flaw.

## 3.4 Formal methods

While addressing runtime-oriented solutions, effort have also been put into formal methods, in the context of the SAMOA project [7]. The authors presented TulaFale [8], a scripting language that formally specifies web service security protocols and analyze their security vulnerability. TulaFale uses pi calculus to specify the interaction among concurrent processes, but also extends pi calculus to include XML syntax and symbolic cryptographic operations for specifying the SOAP message exchange. To specify the construction and verification of SOAP messages, TulaFale uses Prolog-style predicates. The different security goals of a SOAP security specification are specified using assertions:

TulaFale = *PiCalculus +XMLSyntax +predicate + assertion.*

The same authors [7, 9, 8], proposed two new tools in [10]. The language they have proposed is a high level link specification language for specifying intended security goals for SOAP message exchanges among SOAP processors. One of their tools compiles the link specifications to generate WS-Security specifications, then another tool analyzes the generated WS-Security specifications using a theorem prover to verify whether the intended security goals can be achieved by the generated WS-Security specification. This analyzer uses TulaFale script to specify a formal model for a set of SOAP processors and their security checks, and to verify the security goals. According to the authors, the policy-driven web services implementations are susceptible to the usual subtle vulnerabilities of traditional cryptographic protocols. But they stated that their tools can help preventing such vulnerabilities by verifying the policy when it is being compiled from link specifications, and double-check the policy at the time of deployment against their original goals after any modifications.

# 4. LIMITATION OF EXISTING SOLUTIONS

We presented several solutions to counter XML rewriting attacks. These proposals can, however, not detect all range of XML rewriting attack, especially the element wrapping variant. In this section we highlight the main limitation of the existing solutions.

## 4.1 XML Digital Signature

XML represents information using a tree structure. XML Digital Signature allows non-contiguous objects of an XML dataset to be signed separately. The signed object may be referenced using an indirection(URI) by the *Reference* element of the Signature. This indirect referencing does not give any information regarding the actual location of the signed object. Therefore, the signed object can easily be relocated and the Signature value still remains valid. In cases where the location of the data object is important in the interpretation of the semantics associated with the data, this can be exploited by an adversary to gain unauthorized access to protected resources [9]. This is the main limitation of XML Digital Signature.

## 4.2 WS-Security

WS-Security exhibits several flaws:

- ➢ it uses XML Digital signature [3] for signing non-contiguous parts of a SOAP message. Therefore, all the limitations of XML Digital signature are also applicable to WS-Security.
- ➢ it allows multiple security header with the same name to exist in the same SOAP message. This creates a pit fall and can be exploited by an attacker.
- ➢ it does not propose any new security technology. However, it specifies how the existing security technology can be used to secure a SOAP message exchange.
- ➢ it encompasses many other standards like XML Digital Signature, XML Encryption, X.509 certificate, Kerberos ticket... For this reason, the specification became quite complex.

## 4.3 WS-Policy

WS-Policy standard lacks semantics. It provides a mechanism for describing the syntactic aspects of service properties. This introduces a limitation on the policy specification and policy intersection. For example, a provider may specify that its service supports a particular algorithm for the adjustment of data retransmission timeout value and a consumer may define a policy requiring a different algorithm. It might be possible to substitute the required algorithm by the provided algorithm, if they are compatible. However, the current standard does not support this kind of relationship identification. Thus, although it is possible, the interaction between the provider and the consumer will not occur.

## 4.4 WS-Security Policy

Securing a web service using WS-Security Policy is no panacea. It is essentially a domain specific language, which selects cryptographic communications protocols, uses low-level mechanisms that build and check individual security headers. It gives freedom to invent new cryptographic protocols, which are hard to get right, in whatever guise [10].

## 4.5 WSE Policy Advisor

Although WSE Policy Advisor can detect errors that otherwise might be overlooked, it has the following drawbacks:
  ➢ WSE Policy Advisor does not provide any formal guarantees. It only provides a suggestion regarding possible flaws in policy configuration files found by running some queries.
  ➢ WSE Policy advisor shows very poor performance if the policy configuration file becomes complex.
  ➢ The queries run by WS-Policy advisor cannot detect possible existence of signed element reordering attack.

## 4.6 SOAP Account

This approach can successfully detect a wide range of XML rewriting attack. However, it fails to detect all range of XML rewriting attack, as depicted on Figure 5. We can summarize our analysis on the SOAP account approach by the following points:
  ➢ It does not include any mechanism to detect the replay attack. Although *MessageID* or *Timestamp* proposed by the WS-Security can be used for this detection, we should consider the fact that these elements are optional. It is perfectly valid for a SOAP message to not include a *MessageID* or *Timestamp*. In that case, even though the SOAP message contains a SOAP Account element, it is prone to XML Rewriting attack.
  ➢ The approach does not include any mechanism that can uniquely identify the parent of the Signed element. Therefore, an attacker to gain unauthorized access to protected resources, can use this unawareness of SOAP account.
  ➢ The SOAP account itself is prone to XML Rewriting attack. It is specified that the receiver should check for the presence of the SOAP account element after receiving the SOAP message. However, intermediaries can append their own SOAP Account element in the SOAP message. Therefore the number of SOAP account elements in a SOAP message is not fixed. For this reason it is not possible to specify in the security policy, how many SOAP account elements must be present in a SOAP message. The attacker can exploit this problem. He can just cut one of the several SOAP account elements of the SOAP message and paste it into a header element that is not signed and make the role attributes of the header element none and *mustUnderstand* attribute to false, see Figure 5. Then this header element will not be processed by the ultimate recipient or by any of the intermediaries. However during the signature validation the reference of the relocated SOAP Account will be found as it is not removed but only relocated.
  ➢ In SOAP account one of the field is used to keep track of the siblings of the Signed element. However, according to SOAP specification, an intermediary can append its own element in any place of a SOAP message. Therefore this sibling information might change from node to node. It is not specified how this change can be detected by the ultimate receiver at the time of validation of the message.
  ➢ In SOAP account, there is a field that keeps track of the successor of a signed element. According to us this information does not have any role in the process of validation of a SOAP message. XML Digital signature actually signs the Digest value of an XML element. The digest value is calculated on the sub tree rooted at the element that is to be signed. Therefore, if an element is signed all of its children are signed implicitly.

Figure 4 (left column):

```
< Envelope >
 <Header>
  < To>ClientProcessService</ >
  < ReplyTo >
   < Address >RequestClient</ >
  </ReplyTo >
  < MessageID>urn:uuid: 6474</ >
  < Action >""</ >
  <SoapAccount Id="2">
   <NoOfChildOfEnvelope>2</ >
   <NoOfChildOfHeader>6</ >
   <NoOfSignedObject>2</ >
   <ParentOfId1>Envelope</ >
   <ParentOfId2>Header</ >
   <SiblingOfId1>Header</ >
   <SiblingOfId2>To , ReplyTo , MessageID,
Action, Security</ >
  </ >
  <Security>
   <BST id="3">xcvsg....</ >
   <Signature>
    <SignedInfo>
     <Reference URI="#1"/>
     <Reference URI="#2"/>
    <SignatureValue>khakhdskh...</ >
    <KeyInfo>
     <Reference URI="3"/></ ></ ></ >
 <Body Id="1">
  < issueLoanOffer >
   < clientData>
    < clientName>John Abraham</ >
    < clientSSN >RSDNLL80D58Z...
    < employeeData>
     < password>qwerty123</passw...
     < username>Piter Pan</ username></ >
    < loanAmount>100000</ >
    < loanType>2</ ></ ></ >
 </ >
```

Signature's reference to the element with the "Id 2"

Signature's reference to the element with the "Id"

**Figure 4. Message Secured with SOAP Account**

Figure 5 (right column):

```
< Envelope >
 <Header>
  < To>ClientProcessService</ >
  < ReplyTo >
   < Address >RequestClient</ >
  </ReplyTo >
  < MessageID>urn:uuid: 6474</ >
  < Action role="none" mustUnderstand="false">
   <Envelope>
    <Header></ >
    < Body Id="1">
     < issueLoanOffer >
      < clientData>
       < clientName>John Abraham</ >
       < clientSSN >RSDNLL80D58Z139E</ ></ >
      < employeeData>
       < password>qwerty123</password>
       < username>Piter Pan</ username></ >
      < loanAmount id=3>100000</ >
      < loanType>2</ ></ ></ ></ >
  </ >
  <SoapAccount Id="2">
   <NoOfChildOfEnvelope>2</ >
   <NoOfChildOfHeader>6</ >
   <NoOfSignedObject>2</ >
   <ParentOfId1>Envelope</ >
   <ParentOfId2>Header</ >
   <SiblingOfId1>Header</ >
   <SiblingOfId2>To, ReplyTo, MessageID, Action,
Security</ >
  </ >
  <Security>
   <BST id="3">xcvsg....</ >
   <Signature>
    <SignedInfo>
     <Reference URI="#1"/>
     <Reference URI="#2"/>
    <SignatureValue>khakhdskhad...</ >
    <KeyInfo>
     <Reference URI="3"/></ ></ ></ >
 <Body>
  < issueLoanOffer >
   < clientData>
    < clientName>Robert Lewis</ >
    < clientSSN >MNASJDSLEKKR</ ></ >
   < employeeData>
    < password>qwerty123</password>
    < username>Piter Pan</ username></ >
   < loanAmount>500000</ >
   < loanType>2</ ></ ></ >
 </ >
```

The attacker creates a role attribute

The Attacker modified the Body

**Figure 5. XML Rewriting Attack on SOAP Account**

## 4.7 Formal methods

Formal methods in this context are good to analyze and reason about web service protocols. The main limitation resides in their assumption. One assumption, is that a message can be read, written or modified by an attacker if the attacker knows the right key. Otherwise the attacker cannot perform the attack. But we have seen and demonstrated that this is not the case for all sort of attack like in XML rewriting attack. Moreover, a major limitation of the above formalizations is that they do not model insider attacks.

## 5.   OUR RECOMMENDATIONS

We reviewed existing solutions and highlighted their limitations to detect all range of XML rewriting attacks. Here we make some recommendations that needs to be taken into account to ensure integrity of SOAP messages. First, SOAP messages are nothing but XML documents. Therefore we can represent SOAP messages using a tree like structure, see Figure 6. Here let us introduce the following term:
Depth: The depth of an element in a tree represents the length of the path from the root of the tree to the element (for example *Security* element has a depth of 2, *Envelope* depth=0, *Reference* depth=5).

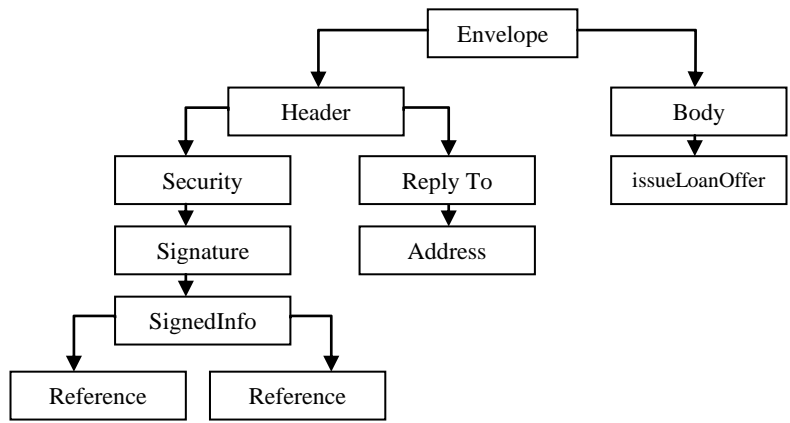

**Figure 6. Tree Representation of a SOAP message**

When looking at Figure 6, we can understand how the Depth can help us. Now it is clear that, if an attacker wraps up a signed element using a fake *Header* then the Depth of the signed element changes.
Our first recommendation is to take into account the Depth information of signed object. But the Depth information itself is not sufficient because it will fail to detect some element wrapping attack.
   We propose to keep information regarding the parent of a signed element. This way if an element is copied and paste the information regarding its parent name will not be the same and the attack will be detected. This is our second recommendation. But parent name information is not reliable, because we can have multiple headers with the same name. We need then a way to uniquely identify the parent of an element. To uniquely identify a parent of an element we rely on WS-Security specification that defines an Id attribute used as follows: *<anyElementwsu:Id = ":::">* ::: *</anyElement>* where *wsu:Id* is of type *xsd:ID*. WS-Security also specifies that two *wsu:Id* within a document cannot have the same value. The usage of this *Id* attribute to uniquely identify a parent of an element is our third recommendation. We have now given 3 recommendations that will help to protect SOAP message against all type of XML rewriting Attacks. We are currently working on the implementation of these recommendations.

## 5.1 Implementation

   Providing a solution to secure SOAP messages against XML rewriting attack, should not increase the effort of developers and must be seamlessly integrated to existing web service stack. Our tests are based on

Apache Axis2 [1] web service implementation. Apache Axis2 provides an execution model as depicted in figure 7. This execution model allows us to interact with the SOAP message before it is processed by the destination application (called *Inflow phase*) but also to work on the response before it reach the client (called *Outflow phase*), by the mean of handlers. Our current implementation direction is to take advantage of this execution model to ensure SOAP message security before it reached client or service side with malformed content or erroneous information due to XML rewriting attack.

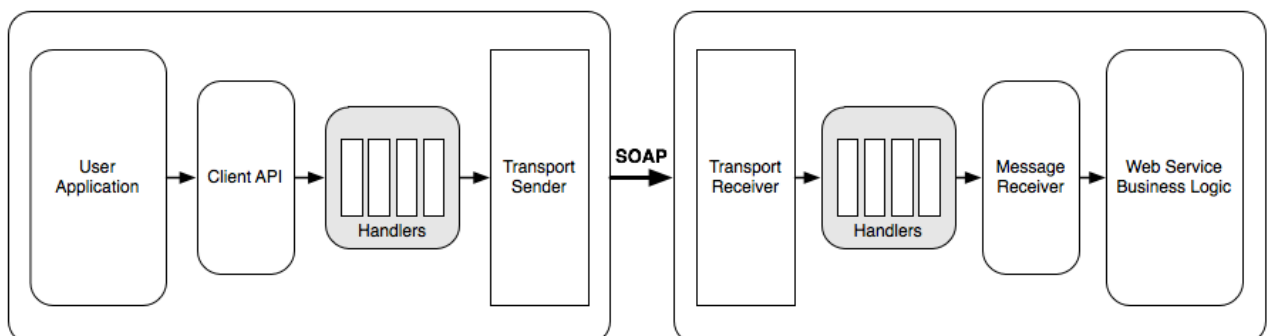

**Figure 7. Apache Axis2 Execution Model**

## 6.  CONCLUSION

In this paper we have presented the nature of the XML rewriting attack, that exist because of the weakness of the XML Signature. We highlighted the fact that while this attack is relatively easy to accomplish, it needs to be taken seriously. Solutions exist to avoid this attack, we showed their limitation to detect all range of attacks and finally presented some ideas to fix this hole. Our future work is to design a solution that relies on the proposed approach and evaluate it.